# Ferroelectric Smectic C Liquid Crystal Phase with Spontaneous Polarization in the Direction of the Director

Hirotsugu Kikuchi[1]*, Hiroya Nishikawa[2], Hiroyuki Matsukizono[1], Shunpei Iino[3], Takeharu Sugiyama[4], Toshio Ishioka[3], Yasushi Okumura[1]*

[1]Kyushu University, Institute for Materials Chemistry and Engineering, 6-1 Kasuga-Koen, Kasuga, Fukuoka, 816-8580, Japan

[2]RIKEN Center for Emergent Matter Science (CEMS), 2-1 Hirosawa, Wako, Saitama, 351-0198, Japan

[3]Kyushu University, Interdisciplinary Graduate School of Engineering Sciences, 6-1 Kasuga-Koen, Kasuga, Fukuoka, 816-8580, Japan

[4]Research Center for Synchrotron Light Applications, Kyushu University, 6-1 Kasuga-Koen, Japan

**Correspondence and requests for materials** should be addressed to
H.K. (kikuchi@cm.kyushu-u.ac.jp) and Y.O. (okumura@cm.kyushu-u.ac.jp**)**

*Abstract*

In our previous study, we demonstrated the existence of an unidentified ferroelectric smectic phase in the low-temperature region of the ferroelectric smectic A phase, where the layer spacing decreases with decreasing temperature. In the present study, we identified the phase by taking 2D X-ray diffraction images of a magnetically oriented sample while allowing it to rotate and constructed a 3D reciprocal space with the sample rotation angle as the third axis for the whole picture of the reciprocal lattice vectors originating from the smectic structure. Consequently, circular diffraction images were obtained when the reciprocal lattice vectors were evenly distributed on the conical surface at a certain inclination angle in the reciprocal space. This result provides clear evidence that the phase in question was smectic C. The polarization properties also showed that the observed smectic C phase had spontaneous polarization in the direction parallel to the director and was identified as ferroelectric smectic C. These results provide a new type of classification for liquid crystalline phases that has been established over many years and is a significant contribution to the basic science of soft matter research.

## 1. Introduction

Ferroelectricity in liquid crystals has attracted the interest of many researchers and engineers and has been actively studied since the pioneering work by Meyer et al.[1] The introduction of chirality into the smectic C phase breaks the mirror symmetry and reduces the symmetry group to $C_2$. Therefore, the spontaneous polarization Ps appears along the $C_2$ axis parallel to the layers of the chiral smectic C phase. Niori et al. discovered that ferroelectricity can manifest in banana-shaped achiral molecules[2]. The diverse molecular arrangements and polarization structures of a family of compounds composed of bent-core molecules subsequently opened a new field of academic research and created a distinctive category in liquid crystal taxonomy. The ferroelectric columnar phase discovered by Miyajima et al. is unique in that it shows spontaneous polarization in the direction of the column axis formed by the umbrella-shaped molecular organization, and the umbrella flips over to exhibit polarization inversion when the electric field is reversed.[3] Ferroelectrics can be classified as proper or improper depending on whether the principal order parameter of the phase transition is polarization. In improper ferroelectrics, polarization emerges through coupling with another order parameter, such as a structural change, which is not directly related to ferroelectricity. The ferroelectric liquid crystals referred to above should be classified as improper ferroelectric phases in the sense that the molecular shape causes structural symmetry breaking and a bias to the rotational motion of the molecules, resulting in an axis where the polarizations do not cancel each other. Conversely, ferroelectric liquid crystals based on a mechanism completely different from the improper type have been discovered in recent years and have attracted considerable attention. Nishikawa et al. identified at least three types of nematic phases in a molecule called DIO (**Figure 1a**); in the lowest-temperature nematic phase, they observed various phenomena in the phase that could only be explained by ferroelectricity, such as a large dielectric constant, polarization inversion behavior, and second harmonic generation, and experimentally demonstrated that the dipole moments are spontaneously aligned in a single direction along the director even though it is a nematic phase.[4] This was the first report of experimental data demonstrating ferroelectricity in a low-molecular-weight nematic phase that exhibits high fluidity similar to a liquid. There are examples of compounds that were initially synthesized without the intention of being ferroelectric and whose ferroelectricity was not confirmed but was later found to be ferroelectric. In a highly polar liquid crystalline compound with fluorine and cyano groups synthesized by Ziobro et al. in 2009,[5] the phase initially determined to be the smectic C phase (SmC) was later identified as the ferroelectric nematic ($N_F$) phase.[6]   In addition, in a compound with a nitro group at the end, synthesized by Mandle et al. in 2017,[7] another nematic phase was present on the low-temperature side of the normal nematic (N) phase, which was initially assigned to the N phase with an anti-parallel molecular arrangement; however, the phase was later identified as the $N_F$ phase.[8] Manabe et al. observed a direct transition from the isotropic phase to the $N_F$ phase in a compound with a core linked by a difluoromethoxy group, which is commonly used in liquid crystal compounds for TFT-driven liquid crystal displays.[9] Recently, ferroelectric nematic phases have attracted considerable interest in soft matter science and device technology.[10] Furthermore, Kikuchi et al. discovered a ferroelectric smectic A ($SmA_F$) phase with spontaneous polarization in the director direction (Figure 1b).[11] These $N_F$ and $SmA_F$ phases can be regarded as proper ferroelectric phases because the polar structure is not stabilized by the molecular steric effect; however, spontaneous polarization is generated by intermolecular dipole-dipole interactions.

In our previous paper, we reported that compound **1** exhibited the $SmA_F$ phase, but an unattributed smectic phase (denoted as $SmX_F$ in the previous paper[11]) was also found in its low-temperature region. Although it was certain that the smectic phase, $SmX_F$, was not smectic A because the layer spacing decreased with decreasing temperature, the phase type could not be clearly identified. In this paper, we report the identification of this smectic phase as ferroelectric smectic C ($SmC_F$) (Figure 1c) based on strong evidence from synchrotron small-angle X-ray diffraction and polarization switching studies. Although the presence of the $SmC_F$ phase has also been suggested as a low-temperature phase for another ferroelectric liquid crystalline material,[12] we have experimentally proven this for compound **1** more clearly. The SmC$_F$ phase discovered in this study is a novel liquid crystal phase that clearly differs from conventional ferroelectric smectic phases, such as the chiral smectic C phase and bent-core smectic phases, in that it shows spontaneous polarization in the direction of the director and an average orientational direction of the molecular long axis.

## 2. Results and Discussion

### 2.1. X-ray diffraction

The chemical structure of liquid crystal compound **1** used in this study is shown in **Figure 1**. In this compound, one of the two fluorine atoms substituted at the 3- and 5-positions of the phenyl group directly connected to the 1,3-dioxane ring of DIO, which exhibits a ferroelectric nematic phase, is replaced by a hydrogen atom. Compound **1** exhibits the following phase transitions on cooling, as shown in previous papers; isotropic phase (I) – 207 ºC - nematic phase (N) -115 ºC - antiferroelectric mesophase (M$_{AF}$, in a previous paper, was referred to as N$_X$ because there was not sufficient confirmation) – 106 ºC – SmA$_F$ - 85 ºC - SmX$_F$, melting point:107 ºC on heating. The 2D X-ray scattering images of compound **1** at different temperatures are illustrated in **Figure 2**. Measurements were performed during the cooling process. 2D WAXD profiles at various temperatures for compound **1** under the magnetic field (*M*-filed, $B = 0.6$ T). The bottom panels show an enlarged one of the red-colored areas of the upper panels. The white double arrow indicates the direction of the *M*-field. The *M*-field was applied in the equatorial direction of the X-ray scattering images, and the liquid crystal director was oriented in this direction. The *d*-spacing calculated from vertically elongated or arc-shaped spots at small angles on the equator was close to the molecular length of the rod-shaped molecule of compound **1**, and the diffraction was attributed to the Bragg reflection from the smectic layer structure. Diffuse scattering in the meridional direction corresponds to the average intermolecular distance perpendicular (lateral) direction to the director. Diffuse scattering indicates that there is no long-range translational order for the molecular arrangement in the smectic layer plane. At temperatures from 120 °C to 130 °C corresponding to the N phase, small-angle spots are relatively weak, but sharp cores are observed at the center. The sharp core of small-angle diffraction became increasingly stronger in the M$_{AF}$ phase at 110 to 115 °C. In the M$_{AF}$ phase, referred to as the N$_X$ phase in the previous paper, the zig-zag (chevron) defect and antiferroelectric behavior were observed by POM and polarization reversal current studies, respectively (see Figure S1 and S2, Supporting Information), suggesting that the M$_{AF}$ phase may be characterized by models, such as the $N_s$[13] and $SmZ_A$[14] phases. When transitioning to the $SmA_F$ phase below 105 °C, the small-angle diffraction peaks became discontinuously stronger and showed clear diffraction owing to the layer structure peculiar to the smectic phase. Sharp second-order diffraction was also observed on the wide-angle side of the first-order diffraction, indicating a high degree of order in the layer structure. Below 80 °C, the small-angle spots split into two; this indicates that the layer planes were tilted with respect to the directors fixed on the equatorial axis by the *M*-field. The meridional scattering remained broad, and the intensity weakened. The phase in this temperature range was described as $SmX_F$ in our previous study because there was insufficient evidence other than that the layer spacing decreased with decreasing temperature, whereas it is very clear from the present experiments that this phase is smectic C. Therefore, we denote this phase as $SmC_F$. The fact that this phase is ferroelectric with spontaneous polarization in the director has been well demonstrated by our previous findings and the latter part of this study. The angle of the spot split widened with decreasing temperature, which is also typical behavior of smectic C. In the case of smectic C, the rotation of the tilting direction of the director with respect to the layer normal, that is, the C director, is a Goldstone mode; therefore, its orientation should be evenly distributed around the layer normal. As the director was fixed in the *M*-field in the current experimental system, the normal vectors of the inclined layer were evenly distributed along the conical surface of the cone with the director as the axis of rotation, and the tilt angle was half the apex angle. Therefore, the distribution of the endpoints of the reciprocal lattice vectors of the smectic C layer is circular in reciprocal space. While 2D XRD shows only two spots per circle as the cross-section with Ewald's reflecting sphere, a whole view of the circle in reciprocal space can be obtained by rotations of the C-director and the circles for the inclined C-layers caused by the rotation of the *M*-field (**Figure 3**a and **3b**). Diffraction images were obtained by rotating the *M*-field around an axis perpendicular to both the *M*-field and incident X-ray in 1-degree increments. The 2D images obtained at each angle were concatenated using the rotation angle as the third axis to form a 3D image (**Figure 3c**). **Figure 3d** shows the first- and second-order diffraction data of compound **1** during the descending temperature process from 130 °C to 70 °C. This image is formed by stacking a total of 31 scattering images observed between −15° and +15° of the rotation angle of the *M*-field with respect to the incident X-ray beam to form a 3D image in reciprocal space and resliced in the axial direction of the rotation angle. Visualization analysis of the intersection of the rotated reciprocal lattice and the Ewald reflection sphere shows ring-shaped diffraction at 70 °C to 80 °C, which clearly proves that the corresponding phase is the $SmC_F$ phase. Furthermore, the point-like diffraction from 85 °C to 110 °C clearly indicates that the corresponding phase is the SmA$_F$ phase. Clear rings appeared in both the first- and second-order diffractions, successfully projecting the entire profile of the reciprocal lattice vectors of the smectic C layers. The fact that the observed rings are continuously connected without interruption and have no particularly large deviation in the intensity distribution indicates that there is no deviation in the orientation of the C-vector around the layer normal, suggesting that there is no significant bias in the Goldstone mode in SmC$_F$ as well as in normal SmC.

**Figure 4**a shows the temperature dependence of the director's tilt angle with respect to the layer normal and layer spacing in the SmA$_F$ and SmC$_F$ temperature ranges. The tilt angle and layer spacing were calculated from the split angle of diffraction and the Bragg angle of diffraction, respectively. In the SmA$_F$ temperature range, the tilt angle is zero, whereas the layer spacing increases with decreasing temperature closer to the SmA$_F$-SmC$_F$ transition temperature. This change may reflect the temperature dependence of the orientational order in the molecular long-axis direction. In the SmC$_F$ phase, the tilt angle increased with decreasing temperature, and synchronously, the layer spacing decreased monotonically. This transition can be regarded as a second-order transition because no discontinuous jump in the tilt angle was observed at the SmA$_F$-SmC$_F$ phase transition temperature. Let us consider the structural changes via the SmA$_F$-SmC$_F$ phase transition using the Madhusudana model, in which a free-rotating molecule is represented as a cylindrical rod with a chain of alternating positive/negative electric charges along the long axes (i.e., a longitudinal surface charge density wave).[15] If each half-wave on the rod has an appropriate amplitude, an N$_F$ phase is induced by the enhanced electrostatic interactions of neighboring rods that are shifted in parallel with respect to each other. Note that such an offset synpolar arrangement was observed by scanning tunneling microscopy of 2D monolayers on an atomically flat crystal surface.[16] Recently, SmC-type long-range correlations within the N$_F$ phase were reported,[17] and the formation of SmC$_F$ structures would naturally be expected if the mutual displacements of the nearest neighbor molecules in the Madhusudana model were considered. Therefore, although there was little molecular displacement level in the SmA$_F$ phase, the displacement level was expected to be optimized to enhance the electrostatic interactions by transitioning to the SmC$_F$ phase (Figure 4b). However, in the SmC$_F$ temperature range, the tilt angle increased continuously with decreasing temperature; that is, the longitudinal displacement of the molecules increased continuously. Ferroelectricity was maintained during this change; this indicates that the Madhusudana model does not necessarily apply to the present system.

## 2.2. Electric polarization properties

To evaluate the ferroelectric behavior of **1**, the polarization reversal current was measured in an in-plane switching (IPS) cell (inset in **Figure 5**a) under a triangular electric field (*E*-field) with a frequency of 0.1 Hz. Figure 5a shows the polarization reversal current (*I*) in the SmC$_F$ regime (*I* in the SmA$_F$ regime is shown in Figure S3, Supporting Information). Although split peaks were observed just after the SmA$_F$–SmC$_F$ phase transition, a single polarization current peak appeared in the SmC$_F$ phase below 83 °C. The electric displacement–electric field (*D*–*E*) hysteresis loops in the SmC$_F$ phase were generated by conversion of the obtained polarization reversal current. For conversion from current to charge, calibration was performed by subtracting the current of the paraelectric/conductive components. Figure 5b shows the obtained *D*–*E* hysteresis loops, which have typical parallelogram shapes observed in common ferroelectric materials. The *D* value and coercive *E*-field (*E*c) in the SmA$_F$–SmC$_F$ regime are shown in Figures 5c and 5d, respectively. With decreasing temperature, the *D* value increased from ca. 3.0 μC cm$^{-2}$ to 3.6 μC cm$^{-2}$. Conversely, the *E*c in the SmA$_F$ phase increased near the SmA$_F$–SmC$_F$ phase transition (up to 1.5 kV cm$^{-2}$), but afterward, the *E*c once dropped to 1.0 kV cm$^{-2}$ and then increased toward the low-temperature side in the SmC$_F$ regime.

This suggests the growth of the correlation length toward the low-temperature side in both the SmA$_F$ and SmC$_F$ phases, thereby increasing the corresponding *E*c. At the SmA$_F$–SmC$_F$ phase transition, *E*c for the newly formed SmC$_F$ should be smaller because the molecules are tilted in a layer. Previously, we demonstrated memorized SHG activity after applying *E*-field during the SmA$_F$ and SmX$_F$ phases. Here, we performed a quantitative evaluation of the remnant polarization after polarization reversal using the positive-up–negative-down (PUND) polarization measurement technique.[18] This technique has been widely utilized for studies on ferroelectric materials because only the polarization reversal current owing to the ferroelectric response can be extracted. Typically, a negative triangular pulse (p) (pre-treatment), two positive (P and U), and two negative (N and D) triangular pulses were subsequentially, and the corresponding currents were recorded (Figure 5f). If the spontaneous polarization is stable during the retention time ($t_r$), the reversal polarization is no longer observed under the subsequent pulse in the U (D) process. In contrast, if the initial polarization is relaxed during $t_r$, the subsequent U (D) pulse restores the polarization. Thus, by subtracting the non-ferroelectric current component in the U (D) process from the entire current component of the P (N) process, the remnant polarization can be estimated. The obtained $D$–$E$ hysteresis loops corresponding to $t_r$ = 1, 10, 100, and 1000 s for the SmC$_F$ phase (70 °C) are shown in the insets in Figure 5e. Notably, no drastic degradation of polarization was observed, even $t_r$ = 1000 s, indicating the high polarization memory function of the SmC$_F$ phase. To evaluate the relaxation kinetics of the polarization for the SmC$_F$ phase, $P(t_r)/P_0$ ($P(t_r)$: remnant polarization at $t_r$; $P_0$: spontaneous polarization) as a function of $t_r$ are plotted (Figure 5f). As illustrated in Figure 5e, only a 6.5% reduction of the initial polarization at $t_r$ = 1000 s ($P(1000)/P_0$ = 0.94) was observed (SmC$_F$: 3.72–3.49 μC cm$^2$).

## 3. Conclusion

We observed a phase in compound **1** in which the layer spacing decreased with decreasing temperature in the low-temperature region of SmA$_F$. Although conventional knowledge of the liquid crystal structure suggests that this phase should be a smectic C phase in which the director is tilted from the layer normal, we considered that further experimental proof was necessary to confirm this; therefore, we called this phase SmX$_F$. In the present study, a full picture of the reciprocal lattice vectors of the layer structure in reciprocal space was captured by rotating the sample and obtaining X-ray diffraction images. Conventional 2D diffraction methods that do not rotate the sample provide only a partial view of the structure. Our 3D reciprocal-space image with sample rotation proved that the low-temperature phase of SmA$_F$ in compound **1** is SmC. The experimental results of the previous study and the precise polarization reversal current measurements performed in this study also demonstrate that this SmC phase is a ferroelectric SmC (SmC$_F$) with spontaneous polarization in the direction of the director. These findings provide a new family of structural classifications of liquid crystals and contribute to the science of liquid crystals and soft matter.

The physical mechanisms underlying the development of ferroelectric nematic phases are not yet fully understood. It is easy to imagine that the interactions between molecular lateral faces are important. In the nematic phase, the dipole moments along the long axis of the molecule are electrostatically repulsive when the molecules are aligned next to each other in the same direction. Therefore, it is preferable for neighboring molecules to be suitably shifted rather than lined up with their heads and tails aligned. It is expected that an appropriate shift will occur depending on the electronic structure of the molecule. The SmA$_F$ that we have previously reported is highly electrostatically unstable when considering only dipolar interactions within a single layer, as the large longitudinal dipole moments of the molecules are aligned in one direction transversely within the layer. The fact that SmA$_F$ is stable suggests the need for a new structural element to develop ferroelectric liquid crystal phases. Considering the experimental fact that there is a smectic short-range order inside the N$_F$ phase, we believe that the stabilization mechanism of the SmA$_F$ phase provides clues for understanding the physical mechanism of the N$_F$ phase. The phenomenon of continuously changing tilt angles in the SmC$_F$s obtained in this study indicates that the longitudinal displacement of the neighboring molecules changes continuously. This suggests that a specific magnitude of displacement does not stabilize ferroelectricity.

## 4. Experimental Section/Methods

*Molecular Syntheses*: The synthetic reaction schemes and molecular characterizations of compound **1** is described in a previous paper. [#9] The phase sequence of compound **1** is as follows: Cry 107 °C (SmC$_F$ 85 °C SmA$_F$ 106 °C) M$_{AF}$ 115 °C N 207 °C Iso. Abbrev.: Cry = crystal, SmC$_F$ = ferroelectric smectic C, SmA$_F$ = ferroelectric smectic A, M$_{AF}$ = antiferroelectric mesophase, N = nematic, Iso = isotropic liquid. This is represented in cooling process, as the SmA$_F$ and SmC$_F$ phases are monotropic in nature.

*Wide-Angle X-Ray Diffraction (WAXD)*: The WAXD measurements were carried out at the Kyushu University beamline, BL06, in the Kyushu Synchrotron Light Research Center (SAGA-LS), Japan using an X-ray wavelength of 1.8 Å (6.89 keV) and a camera length of 179.809 mm. Images were taken in transmission geometry using a two-dimensional pixel-array detector (PILATUS-300K, DECTRIS (4). An aluminum sample folder with a thickness of 2 mm and a hole diameter of 2.4 mm held a 2 mm thick liquid crystal sample by surface tension without the use of a Kapton window (Figure S4, Supporting Information). A *M*-field of 0.6 T was applied to the sample in the equatorial direction by a magnetic circuit consisting of samarium-cobalt magnets with excellent heat resistance and iron cores. The entire magnetic circuit was housed in an aluminum temperature stage box to enhance the accuracy of the temperature control. The temperature of the stage box was controlled by a digital temperature controller (TC-1000A, AS ONE Corporation). The stage box was mounted on a 6-axis control stage (Hexapods, Physik Instrumente), and rotation around an axis perpendicular to the incident X-ray and *M*-field was controlled by LabVIEW, as shown in Figure S5, Supporting Information.

*Polarization Reversal Current and D–E Hysteresis*: *D–E* hysteresis measurements were performed in the temperature range of the SmA$_F$ and SmC$_F$ phase under a triangular-wave electric field (0.1 Hz, 400 V) using an FCE system (TOYO Corporation), which equipped with an arbitrary waveform generator (2411B), an IV/QV amplifier (model 6252) and a simultaneous A/D USB device (DT9832). We used a IPS cell (thickness: 10 μm, electrode distance: 1 mm, electrode length: 18 mm, E.H.C.) with anti-parallel rubbing condition. In this experiment, the sample temperature was controlled using the INSTEC model mK2000 temperature controller and a liquid nitrogen cooling system pump (LN2-P/LN2-D2, INSTEC).

**Author Contributions**

H.K. supervised the project and wrote the manuscript. H.N. suggested analysis of ferroelectricity and performed polarization reversal current, $D$–$E$ hysteresis measurements, and data analysis. H.N. performed POM studies and characterize the antiferroelectric phase. H.N. and Y.O. contributed to the graphical visualization of the data. H.N. and Y.O. partially wrote the manuscript. H.M. synthesized liquid crystals. T.S. was the beamline manager of synchrotron XRD and T.S. and I.T. setup a device to rotate samples using Hexapods. Y.O. proposed the XRD measurement with a rotating $M$-field, fabricated the temperature stage with a magnetic circuit, and analyzed the XRD data. S.I, Y.O and T.S. performed synchrotron XRD measurements. S.I. analyzed the XRD data. All authors discussed the results and contributed to the manuscript.

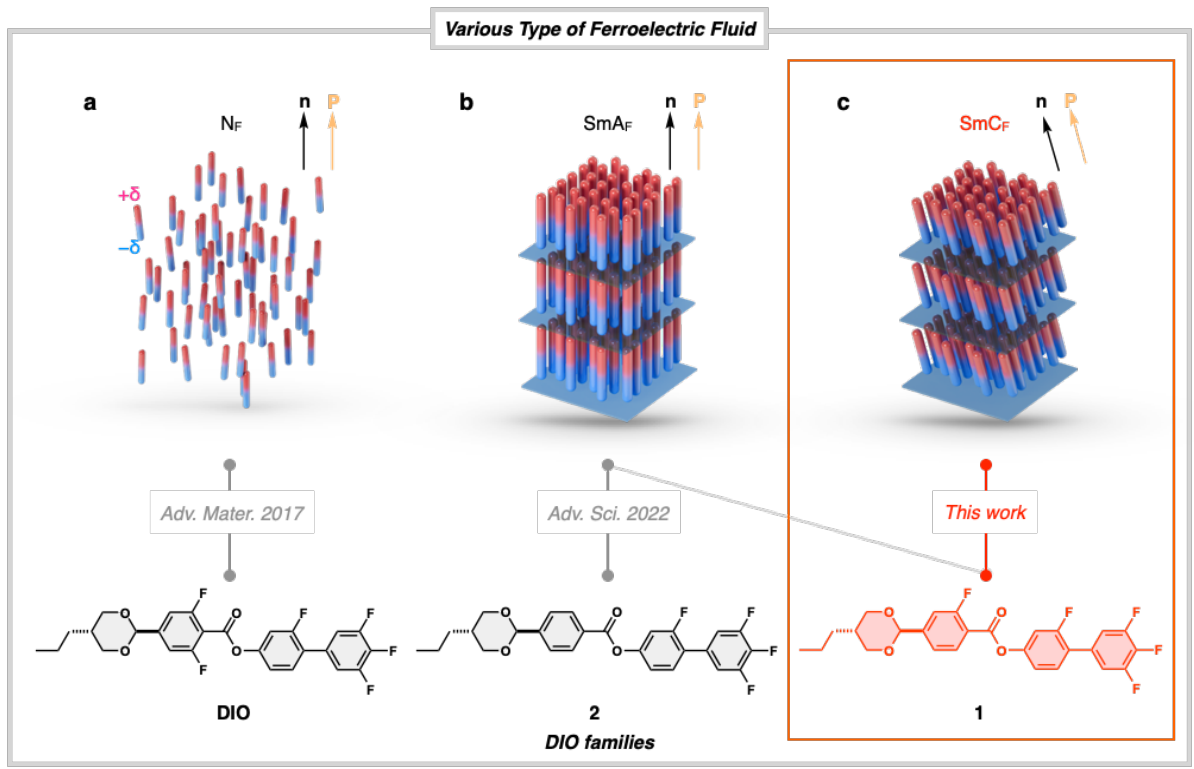

**Figure 1**. Schematic illustrations of a ferroelectric fluid hierarchy categorized into three phases: (a) ferroelectric nematic ($N_F$), (b) ferroelectric smectic A ($SmA_F$), and (c) ferroelectric smectic C ($SmC_F$) phases. An archetypal molecule (DIO) shows the $N_F$ phase. Compounds **1** and **2**, with modification of the DIO's structure, exhibit $SmA_F$/ $SmC_F$ and $SmA_F$ phases, respectively.

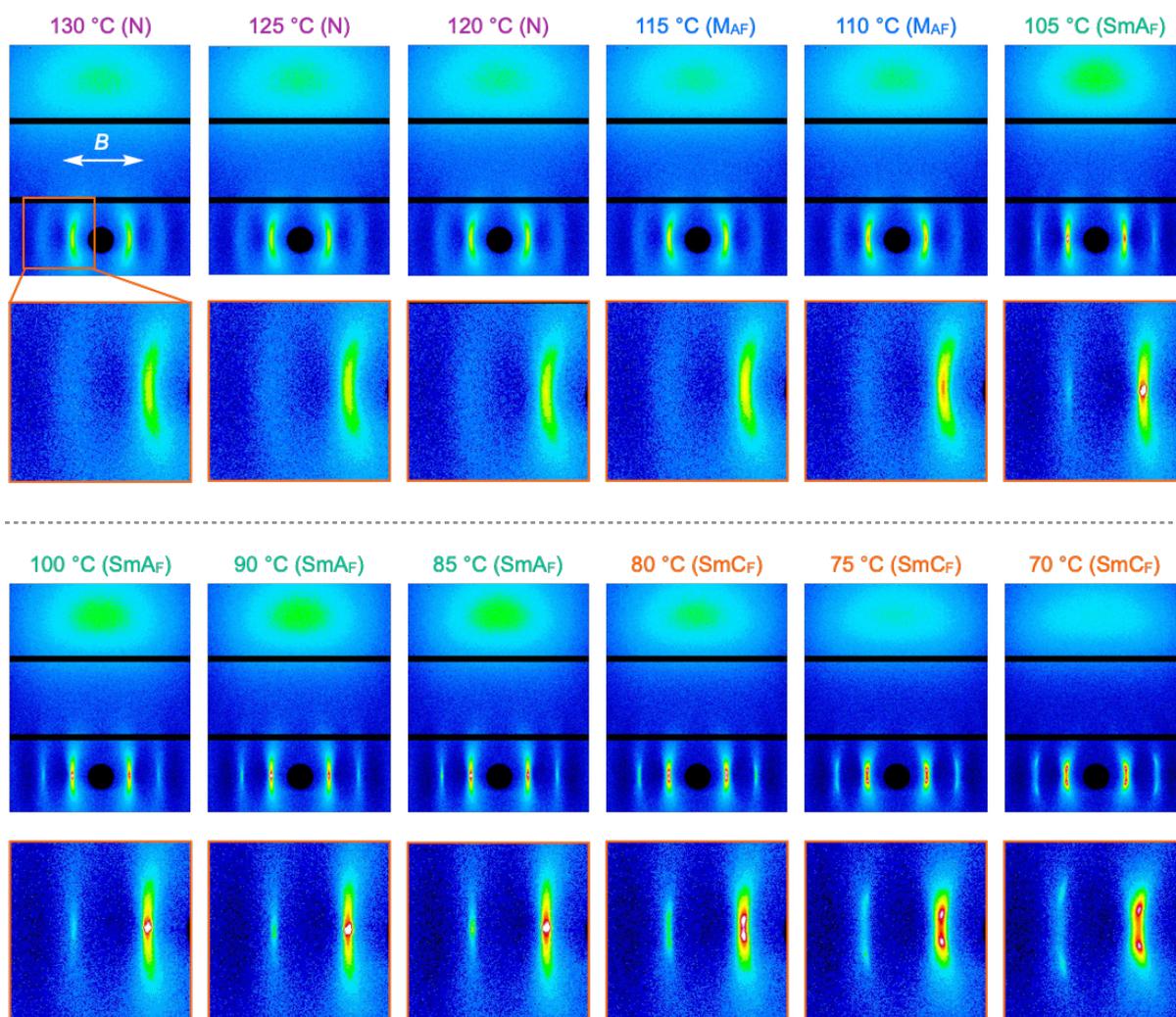

**Figure 2**. 2D WAXD profiles in various temperatures for compound **1** under the *M*-filed (0.6 T). The bottom panels denote an enlarged one of the red-colored areas in the upper panels. A white double arrow indicates the direction of the *M*-field.

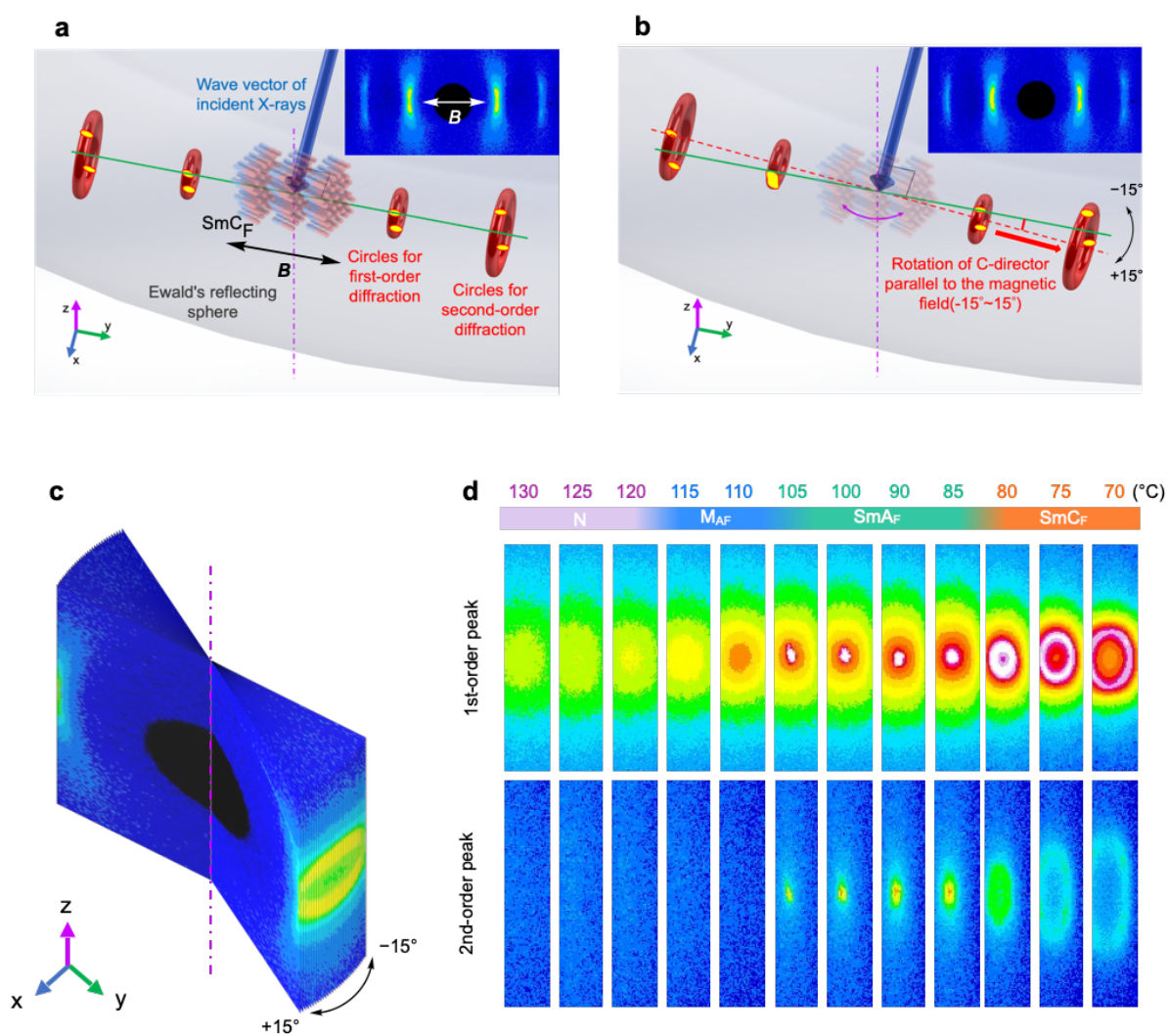

**Figure 3**. Phase characterization based on X-ray diffraction for compound **1**. a–c) The schematic illustration of geometries for characterization of a typical smectic C phase. An example of a 2D XRD pattern in the SmC$_F$ phase is inserted in the upper right corner in the panels (a) and (b). a) Blue arrows and gray spheres represent the wave vector and Ewald's reflecting sphere of incident X-rays, respectively. Small and large red circles represent the 1st- and 2nd-diffraction of C-layers in the reciprocal space. The yellow region represents the intersection of Ewald's reflecting sphere and the diffraction of the C-layers. Projecting the regions onto the Y–Z plane from the center of Ewald's reflecting sphere yields the 2D diffraction image of the C-layer. b) Changes induced by the rotation of the C-layer diffraction due to the rotation of the *M*-field. c) A diffraction circle formed by C-layers in wave space obtained by cutting stacked WAXD images in the X–Z plane. d) Generated 2D XRD patterns in various phases.

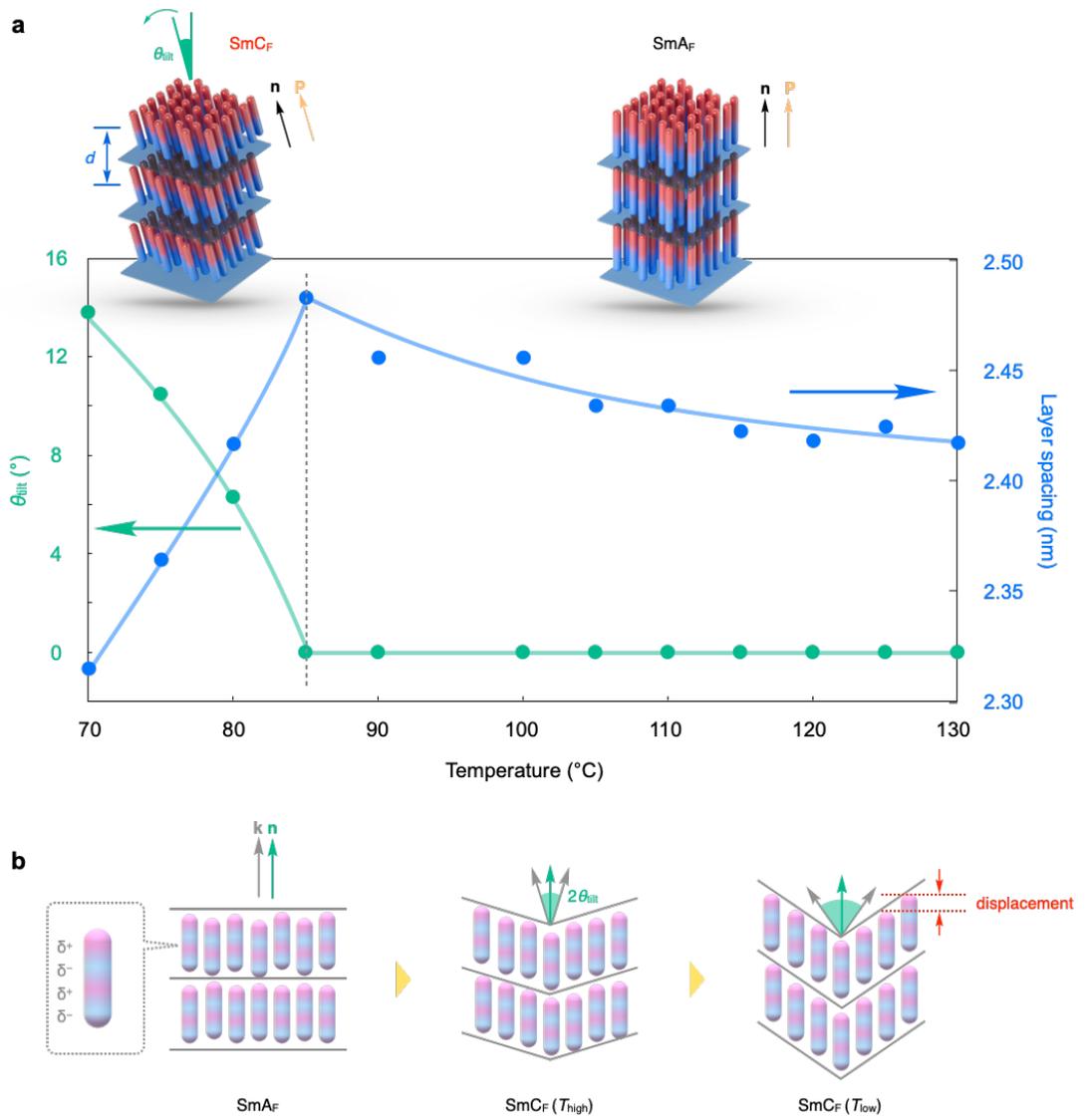

**Figure 4**. a) Temperature dependence of tilt angle ($\theta_{tilt}$) and layer spacing (*d*) in the SmA$_F$ and SmC$_F$ phases. b) Schematic illustration of structure change from SmA$_F$ to SmC$_F$ phases. Here, k and n represent normal and director vectors, respectively.

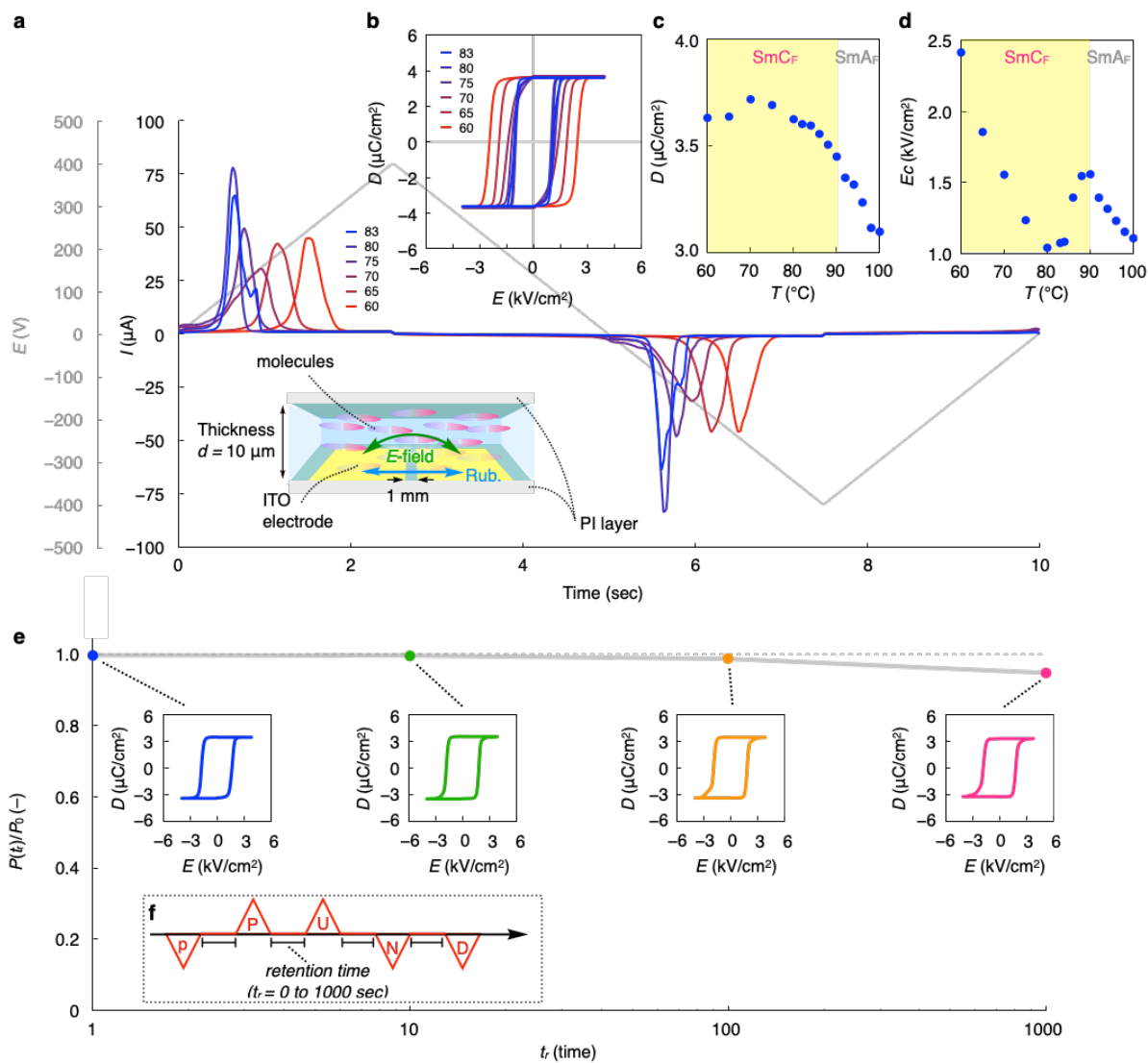

**Figure 5**. Polarization behavior of SmC$_F$ phase for compound **1**. a) The reversal polarization current in various temperatures for the SmC$_F$ phase (triangular waveform, $f$ = 0.1 Hz, $E$ = 400 V). The inset represents the schematic illustration of the IPS cell for $D$–$E$ hysteresis measurement. b) the converted $D$–$E$ hysteresis. $D$ vs $T$ (c) and $E$c vs $T$ (d) in the SmA$_F$ and SmC$_F$ phases. e) The illustration of the PUND waveform. The retention time was set to be 1, 10, 100, 1000 sec. f) The relaxation kinetics of polarization for the SmC$_F$ phase at 70 °C ($f$ = 0.1 Hz, $E$ = 400 V).

**Acknowledgements**

This work was supported by the Grant-in-Aid for Scientific Research (A) JSPS KAKENHI Grant Number JP 23H00303 and JP23K17366 from the Japan Society for the Promotion of Science (JSPS), MEXT Project "Integrated Research Consortium on Chemical Sciences (IRCCS)," Dynamic Alliance for Open Innovation Bridging Human, Environment and Materials from the Ministry of Education, Culture, Sports, Science and Technology, Japan (MEXT), the Co- operative Research Program of "Network Joint Research Center for Materials and Devices". This work was partially supported by JSPS KAKENHI (JP22K14594; H.N.), RIKEN Special Postdoctoral Researchers (SPDR) fellowship (H.N.), RIKEN Incentive Research Projects (FY2024: H.N.). The synchrotron XRD experiments were performed with the approval of Kyushu University Beamline (Proposal Nos. 2022IIK001). We are grateful to Dr. F. Araoka (RIKEN, CEMS) for allowing us to use a ferroelectricity evaluation system.

# Supporting Information

**Ferroelectric Smectic C Liquid Crystal Phase with Spontaneous Polarization in the Direction of the Director**

*Hirotsugu Kikuchi\*, Hiroya Nishikawa, Hiroyuki Matsukizono, Shunpei Iino, Takeharu Sugiyama, Toshio Ishioka, Yasushi Okumura\**

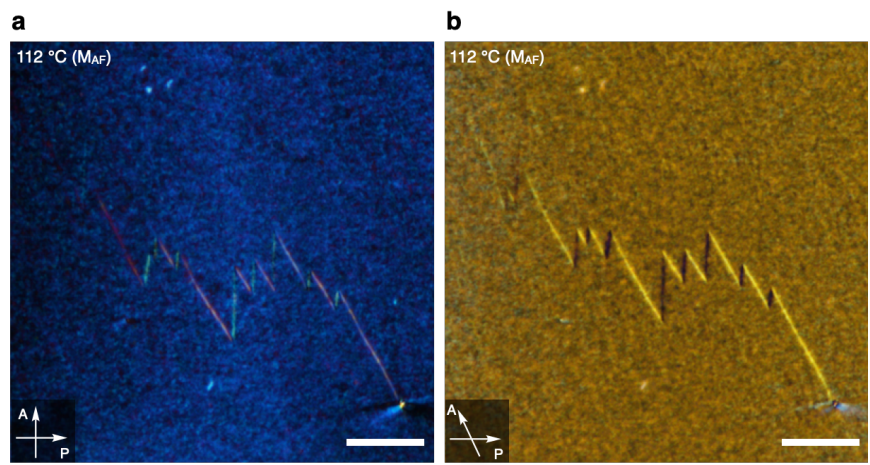

**Figure S1**. POM images for the M$_{AF}$ phase (112 °C) in a polyimide cell (4 μm) under crossed-(a) and decrossed-(b) polarizers. Scale bar: 200 μm.

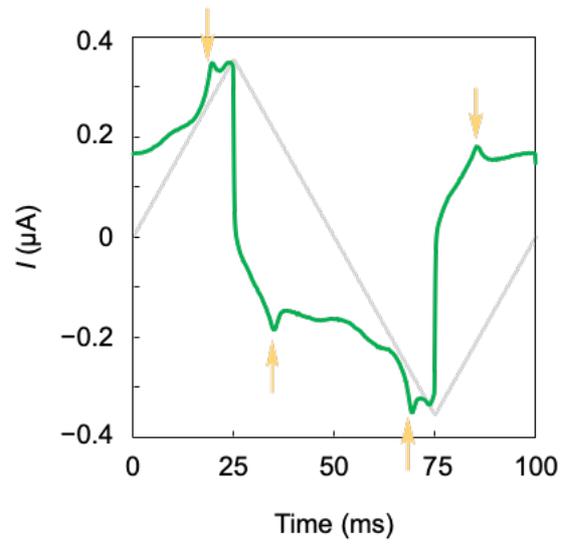

**Figure S2**. Polarization reversal current measurement in the M$_{AF}$ phase (112 °C) under the *E*-field with a triangular voltage wave (400 V, 10 Hz). The orange arrows denote the polarization reversal current peaks.

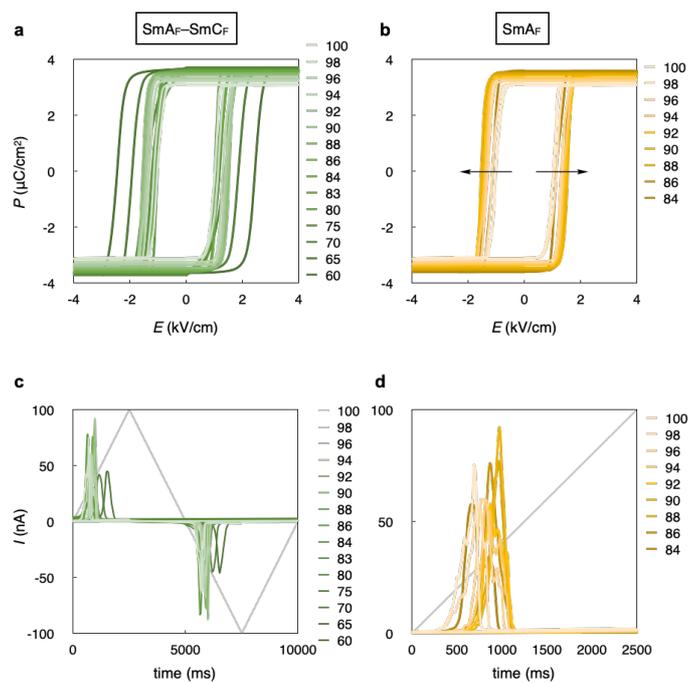

**Figure S3**. *D–E* hysteresis loops (a,b) and polarization reversal current profiles (c,d) for compound **1**. a,c ) SmA$_F$ and SmC$_F$ regimes, b,d) SmA$_F$ regime. All data were recorded under the *E*-field with a triangular voltage wave (400 V, 0.1 Hz).

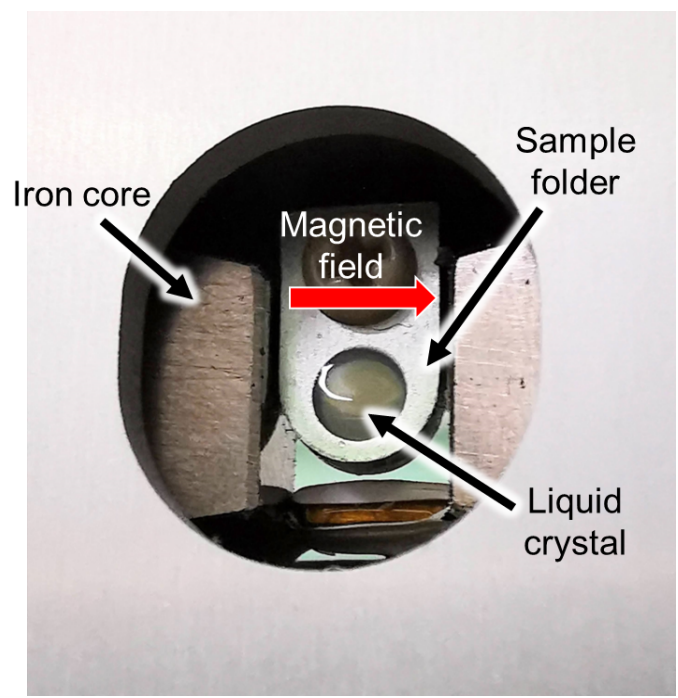

**Figure S4**. (a) Sample folder for liquid crystals mounted on the temperature stage box with magnetic circuit.

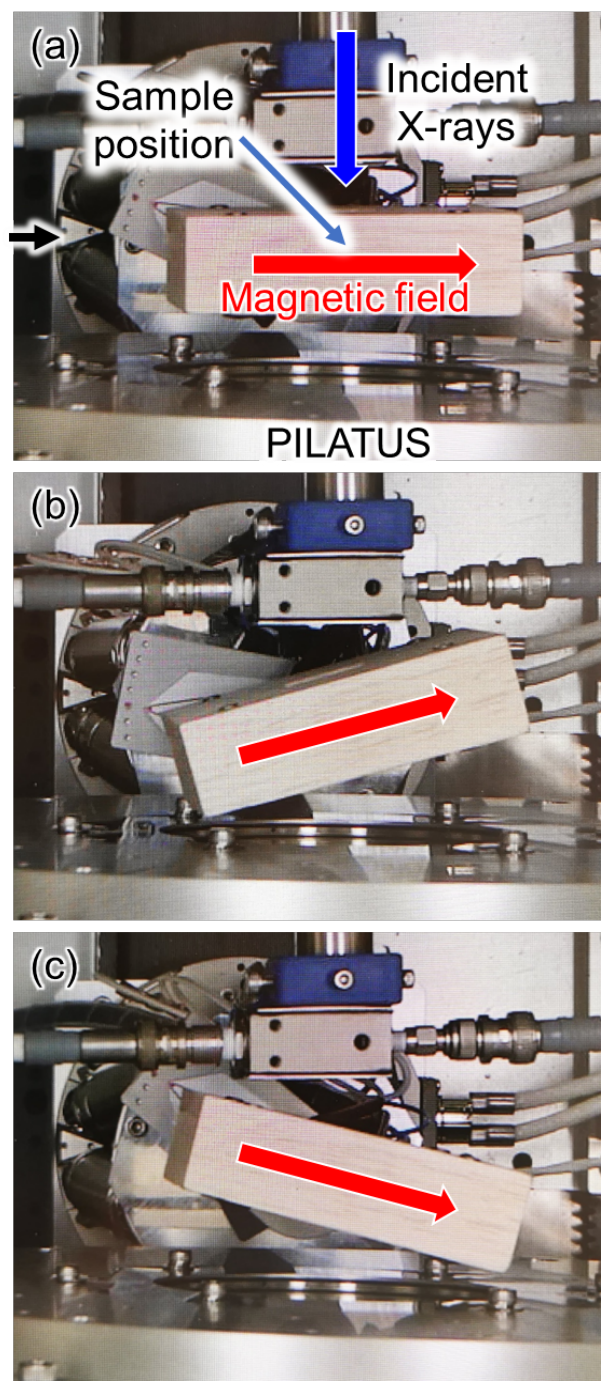

**Figure S5**. Rotation of the *M*-field application temperature stage box attached to the hexapods. Rotation angle = 0° (a), −15° (b), 15° (c).